\documentclass[aps,prb,superscriptaddress,twocolumn,nofootinbib,floatfix]{revtex4-2}

\usepackage{amssymb}
\usepackage{amsmath}
\usepackage{appendix}
\usepackage{times}
\usepackage{graphicx}
\usepackage{subeqnarray}
\usepackage{bbold}
\usepackage{enumerate}
\usepackage{xcolor}
\usepackage{bm}
\usepackage{microtype} 
\usepackage[colorlinks=true, letterpaper=true, pdfstartview=FitV,  linkcolor=blue, citecolor=blue, urlcolor=blue]{hyperref}

\usepackage[normalem]{ulem}

\setcitestyle{super,open={},close={}}

\makeatletter
\renewcommand\@biblabel[1]{#1.}
\makeatother

\begin{document}
\title{Hybrid-parity sliding multiferroics}

\author{Zhenzhou Guo}
\address{Institute for Superconducting and Electronic Materials, Faculty of Engineering and Information Sciences, University of Wollongong, Wollongong 2500, Australia.}

\author{Jiangtao Yu}
\address{Institute for Superconducting and Electronic Materials, Faculty of Engineering and Information Sciences, University of Wollongong, Wollongong 2500, Australia.}

\author{Shibo Fang}\email{shibo\_fang@sutd.edu.sg}
\affiliation{Science, Mathematics and Technology (SMT) Cluster, Singapore University of Technology and Design, Singapore 487372.}

\author{Jin Cao}
\email{caojin.phy@gmail.com}
\affiliation{Research Laboratory for Quantum Materials, Department of Applied Physics, The Hong Kong Polytechnic University, Kowloon, Hong Kong SAR, China.}

\author{Xiaodong Zhou}
\address{Institute of Quantum Materials and Devices, Tiangong University, Tianjin 300387, China.}

\author{Yee Sin Ang}
\affiliation{Science, Mathematics and Technology (SMT) Cluster, Singapore University of Technology and Design, Singapore 487372.}

\author{Wenhong Wang}
\address{Institute of Quantum Materials and Devices, Tiangong University, Tianjin 300387, China.}

\author{Zhenxiang Cheng}\email{cheng@uow.edu.au}
\address{Institute for Superconducting and Electronic Materials, Faculty of Engineering and Information Sciences, University of Wollongong, Wollongong 2500, Australia.}

\author{Xiaotian Wang}\email{xiaotianw@uow.edu.au}
\address{Institute for Superconducting and Electronic Materials, Faculty of Engineering and Information Sciences, University of Wollongong, Wollongong 2500, Australia.}

\begin{abstract}
In this work, we introduce a class of hybrid-parity sliding multiferroics in which the spontaneous ferroelectric polarization is coupled to certain nonrelativistic spin splitting components through interlayer sliding, allowing these components to be reversibly switched in an electrical way. Symmetry analysis identifies coplanar magnets as natural platforms for realizing this form of sliding multiferroicity. First-principles calculations establish bilayer VBr$_2$ as a representative example, demonstrating the coupled reversal of the out-of-plane ferroelectric polarization and the signs of both even- and odd-parity nonrelativistic spin splitting components via an interlayer-sliding pathway. The signs of these nonrelativistic spin splitting components are locked to the sliding-switchable ferroelectric polarization and encoded in the spin-current responses, providing a signature of the coupled ferroic switching. Our findings expand the scope of sliding multiferroics and the functionality of sliding ferroelectrics for low-energy, nonvolatile logic devices.
\end{abstract}

\maketitle

Sliding ferroelectricity in two-dimensional van der Waals layered materials has emerged as an important route to realizing atomic-scale ferroelectricity, with systems usually constructed by stacking high-symmetry nonpolar layers into polar configurations whose ferroelectric polarization can be reversed by interlayer sliding\cite{liBinaryCompoundBilayer2017,wu100YearsFerroelectricity2021,wuSlidingFerroelectricity2D2021,jiGeneralTheoryBilayer2023,liVanWaalsFerroelectrics2024}. An expanding range of van der Waals systems, including boron nitride\cite{yasudaStackingengineeredFerroelectricityBilayer2021a,viznersternInterfacialFerroelectricityVan2021,yeoPolytypeSwitchingSuperlubricant2025} and transition metal dichalcogenides\cite{feiFerroelectricSwitchingTwodimensional2018,debCumulativePolarizationConductive2022,westonInterfacialFerroelectricityMarginally2022b,wangInterfacialFerroelectricityRhombohedralstacked2022a,rogeeFerroelectricityUntwistedHeterobilayers2022a}, exhibiting sliding ferroelectricity, has been reported experimentally. In these systems, the low switching barrier, arising from weak interlayer coupling, enables ultrafast polarization reversal with low energy cost\cite{zhongTheoreticalDesignsLowbarrier2023}. Meanwhile, the interlayer-sliding-mediated ferroelectric switching suppresses defect accumulation and domain-wall pinning\cite{bianDevelopingFatigueresistantFerroelectrics2024}, ensuring fatigue-free and high-endurance operation\cite{yasudaUltrafastHighenduranceMemory2024,fanTailoredSlidingFerroelectricity2025}. Beyond polarization switching, sliding ferroelectricity provides a versatile means of electrically tuning band topology\cite{yangSlidingFerroelectricsInduced2025}, superconductivity\cite{jindalCoupledFerroelectricitySuperconductivity2023}, and magnetism\cite{bennettStackingEngineeredFerroelectricityMultiferroic2024,guoSlidingFerroelectricMetal2025}.

Building on these advances, sliding ferroelectricity offers a bottom-up approach based on van der Waals assembly for engineering ultrathin ferroelectricity in unconventional magnets that feature momentum-dependent nonrelativistic spin splitting (NSS) despite compensated magnetization~\cite{zhuSlidingFerroelectricControl2025b,sunProposingAltermagneticFerroelectricTypeIII2025,yanIonicSlidingFerroelectricity2025,sunUnifiedSymmetryFramework2026,sunSymmetrylockedSixstateControl2026a}. Previous efforts, however, have primarily focused on electrically switchable even-parity NSS in collinear altermagnets. More broadly, recent studies suggest that in compensated noncollinear magnets, the NSS can exhibit odd-parity or even hybrid-parity~\cite{hellenesPwaveMagnets2024a,yamadaMetallicPwaveMagnet2025,luoSpinGroupSymmetry2026a,priessnitzFerroelectric$p$waveMagnets2026a,luoUnconventionalMagnetismSymmetry2026b}. A typical example is provided by the $p$-wave magnets, and a recent experiment has demonstrated ferroelectric switching of $p$-wave NSS on the noncollinear type-II multiferroic NiI$_2$~\cite{songElectricalSwitchingPwave2025}, highlighting ferroelectric polarization as a nonvolatile control parameter for unconventional spin splitting. On the other hand, in hybrid-parity magnets, the spin components possess well-defined but different parities. This provides a unique platform for studying both odd- and even-parity-related physical effects and expands the playground of unconventional magnets. However, it remains unclear how to electrically control hybrid-parity NSS.

Here, we explore the sliding ferroelectric manipulation of NSS beyond collinear altermagnets by introducing the concept of hybrid-parity sliding multiferroicity. In such a multiferroic, the spontaneous ferroelectric polarization is coupled to certain NSS components through interlayer sliding, such that electrically reversing the polarization switches the signs of these components. We establish the symmetry requirements for this form of multiferroicity and identify coplanar magnets as natural candidate platforms for its realization. Building on these symmetry considerations, we propose bilayer VBr$_2$ with coplanar magnetic order as a representative example. First-principles calculations reveal that it exhibits both even- and odd-parity NSS components, with the signs of the odd-parity $x$ component and the even-parity $z$ component reversing together with the ferroelectric polarization via interlayer sliding. The signs of these switchable components are encoded in the spin-current responses, thereby providing a signature of the polarization-coupled hybrid-parity NSS switching.

\vspace{0.3cm}
\noindent{\Large \bf Results}

\vspace{0.05cm}	
\noindent{\bf The concept of hybrid-parity sliding multiferroics}

\begin{figure*}[t]
\centering
\includegraphics[width=1.53\columnwidth]{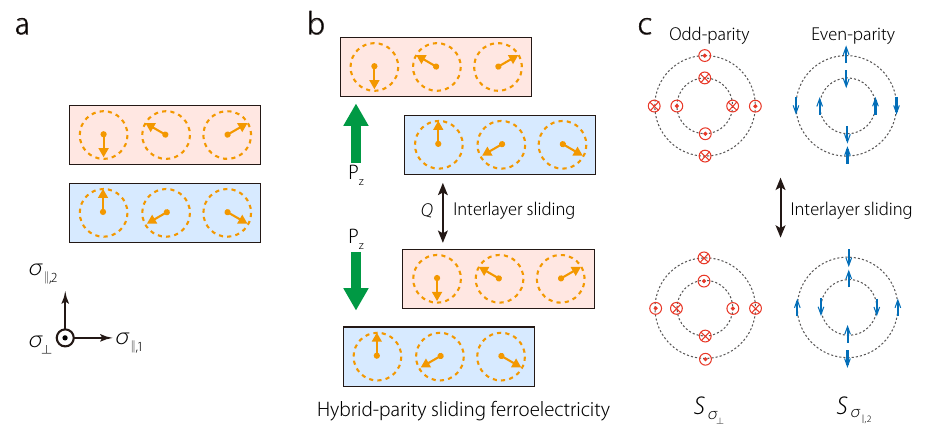}
\caption{\textbf{Schematic diagram of hybrid-parity sliding multiferroics.}
\textbf{a}, A bilayer coplanar magnet, constructed from a monolayer, exhibits the symmetry $Q=\{ C_{2\sigma_{\Vert,1}}{\parallel}\tilde{M}_{z} \} $. $\sigma_{\bot}$ is the direction perpendicular to the magnetic moments, while $\sigma_{\Vert,1}$ and $\sigma_{\Vert,2}$ denote the in-plane directions. 
\textbf{b}, A spontaneous ferroelectric polarization $P_z$ induced by interlayer sliding in \textbf{a} breaks the $Q$ symmetry. This results in two stacking states connected by interlayer sliding, which are related by $Q$ and exhibit opposite ferroelectric polarizations. 
\textbf{c}, The corresponding switchable NSS in \textbf{b}. The NSS exhibits both even- and odd-parity components due to the coplanar magnetic ordering. Meanwhile, the even-parity component $\boldsymbol{S}_{\sigma_{\Vert,2}}$ and the odd-parity component $\boldsymbol{S}_{\sigma_{\bot}}$ are switchable via interlayer sliding, enabling a coupling between the ferroelectricity and the hybrid-parity NSS. 
\label{fig1}}
\end{figure*}


\noindent
In the description of a magnetic space group,  rotations in spin space are locked to the corresponding rotations in real space. While this is strictly correct, the physics within the energy window of interest may exhibit more symmetry\cite{brinkmanSpaceGroupTheory1966}. In unconventional magnets, a central concept is NSS, which encodes the symmetry and configuration of the underlying magnetic order~\cite{Wu2007Fermi,Hayami2019Momentum,Yuan2020Giant,Smejkal2022conventional,guoSpinPolarizedAntiferromagnetsSpintronics2025a,Liu2025Different,jungwirthSymmetryMicroscopySpectroscopy2026}. As a nonrelativistic effect, NSS is subject to the constraints of spin space group (SSG) symmetry~\cite{liuSpinGroupSymmetryMagnetic2022,chenEnumerationRepresentationTheory2024,xiaoSpinSpaceGroups2024b,Jiang2024Enumeration}, in which rotations in spin and real space can be defined independently.
An SSG operation can be expressed as $\{ XU{\parallel}\tilde{R}\}$, where $X$ is the identity or time-reversal symmetry $T$, $U$ is a rotation operation acting on spin, and $\tilde{R} = \{R|\boldsymbol{\tau}\}$ is a space group symmetry acting on the lattice, with $\boldsymbol{\tau}$ being either trivial or nontrivial. A general SSG takes the structure $\mathcal{G}_{S}=\mathcal{S}\times \mathcal{G}$. Here, $\mathcal{S}$ is the spin-only group, defined as the set of operations in $\mathcal{G}_{S}$ that take the form $\{ XU{\parallel}1|\boldsymbol{0}\} $. The spin-only group is determined solely by whether the real-space magnetic structure is collinear, coplanar, or noncoplanar. The collinear case is irrelevant to hybrid-parity NSS and is therefore ignored. For noncoplanar magnets, $\mathcal{S}$ contains only the identity operation. For coplanar cases, $\mathcal{S}$ is generated by $\{ TC_{2\sigma_{\bot}}{\parallel}1|\boldsymbol{0}\} $, where $\sigma_{\bot}$ denotes the direction perpendicular to the magnetic moments, and $C_{2\sigma_{\bot}}$ denotes a twofold rotation along $\sigma_{\bot}$. $\mathcal{G}$ is the quotient SSG as suggested by the structure of SSG, where the operation could have nontrivial spin rotation.

he SSG approach has been widely used to investigate the coupling between NSS and ferroelectricity\cite{Gu2025Ferroelectric,sunProposingAltermagneticFerroelectricTypeIII2025,zhuSlidingFerroelectricControl2025b,sunUnifiedSymmetryFramework2026,zhaoMultiferroicPhaseTransition2026,yanIonicSlidingFerroelectricity2025,sunSymmetrylockedSixstateControl2026a,Duan2025Antiferroelectric,xieChiralAltermagneticMagnetoelectrics}.
To realize a hybrid-parity multiferroic with interlayer sliding switchable functionality, the following requirements must be fulfilled: (i) The SSG allows well-defined hybrid parity of the NSS. (ii) The quotient SSG is compatible with ferroelectricity. (iii) The SSG operation that connects nonequivalent ferroelectric states simultaneously reverses ferroelectric polarization and certain components of the NSS, therefore establishing a coupling between them.

For a well-defined parity of the spin expectation, the system should possess an SSG operation that connects $\pm\boldsymbol{k}$. For a two-dimensional system, the possible operations include $\{ U{\parallel}\tilde{P}\} $, $\{ U{\parallel}\tilde{C}_{2z}\} $, $\{ TU{\parallel}\tilde{1}\} $ or $\{ TU{\parallel}\tilde{M}_z\} $, where we have omitted the translational part for simplicity, $P$ denotes inversion symmetry, and $U$ can be the identity or twofold rotation. Among them, $\{ U{\parallel}\tilde{P}\} $ rules out any ferroelectricity, which does not meet our requirement. $\{ TU{\parallel}\tilde{1}\} $, $\{ U{\parallel}\tilde{C}_{2z}\} $ and $\{ TU{\parallel}\tilde{M}_z\} $ are compatible with any polarization, polarization along $z$, and polarization in the $xy$ plane, respectively. Taking $\{ TU{\parallel}\tilde{1}\} $ as an example, it enforces pure odd-parity NSS when $U$ is the identity, and enforces hybrid-parity NSS when $U$ is a twofold rotation. 

It is worth noting that coplanar magnets serve as natural candidates for realizing hybrid-parity sliding multiferroics, since their spin-only group is generated by $\{ TC_{2\sigma_{\bot}}{\parallel}1|\boldsymbol{0}\} $. It requires that 
\begin{equation}
\boldsymbol{S}_{\sigma_{\bot}}\left(\boldsymbol{k}\right)=-\boldsymbol{S}_{\sigma_{\bot}}\left(-\boldsymbol{k}\right),\quad\boldsymbol{S}_{\sigma_{\Vert}}\left(\boldsymbol{k}\right)=\boldsymbol{S}_{\sigma_{\Vert}}\left(-\boldsymbol{k}\right),\label{eq1}
\end{equation}
where $\boldsymbol{S}_{\sigma_{\bot}}$ denotes the perpendicular component along $\sigma_{\bot}$ and $\boldsymbol{S}_{\sigma_{\Vert}}$ denotes the remaining in-plane component. Therefore, the spin-only group of the coplanar magnets requires odd parity of NSS of the perpendicular component, and even parity of NSS of the in-plane component. We note that the detailed form of the electric polarization is subject to the constraint of the quotient SSG, and the spin-only group itself does not impose a constraint on the polarization. 

The SSG operation $Q$ that connects nonequivalent ferroelectric states should leave $\boldsymbol{k}$ invariant while reversing certain NSS components. Meanwhile, it should allow reversal of electric polarization via interlayer sliding. The possible operations are $\{ U{\parallel}\tilde{M}_{z}\} $ for $U$ being a twofold rotation and $\{ TU{\parallel}\tilde{C}_{2z}\} $ for $U$ being either the identity or twofold rotation. 

As an example, consider a bilayer coplanar magnet as shown in Fig.~\ref{fig1}a, which exhibits the symmetry $Q=\{ C_{2\sigma_{\Vert,1}}{\parallel}\tilde{M}_{z}\} $. If a spontaneous ferroelectric polarization along $z$ arises from interlayer sliding, the $Q$ symmetry is broken, giving rise to two degenerate stacking states, as shown in Fig.~\ref{fig1}b. The two stacking states are related by $Q$, and therefore exhibit opposite ferroelectric polarizations. Since the interlayer sliding preserves the coplanar magnetic configuration, the ferroelectric states maintain coplanar magnetic ordering. Therefore, the ferroelectric states allow hybrid-parity NSS given by Eq.~(\ref{eq1}). Upon ferroelectric switching, the $Q$ symmetry requires the switching of both the even-parity component $\boldsymbol{S}_{\sigma_{\Vert,2}}$ and the odd-parity component $\boldsymbol{S}_{\sigma_{\bot}}$, as shown in Fig.~\ref{fig1}c.

We note that the SSG-based symmetry results presented above are applicable when spin-orbit coupling is relatively weak within the energy window of interest. We also note that in this regime, the coupling between ferroelectricity and NSS is primarily mediated by exchange interactions, which can be well captured within the SSG symmetry framework.

\begin{figure}[t]
\centering
\includegraphics[width=\columnwidth]{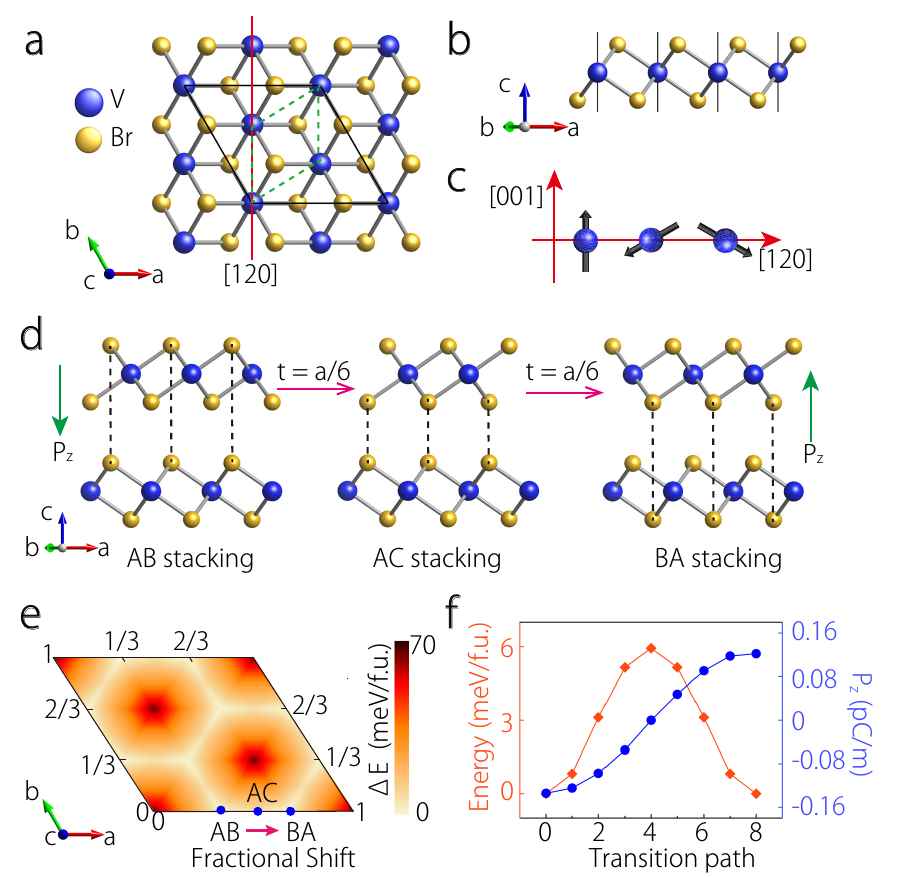}
\caption{\textbf{Crystal structure and sliding ferroelectric switching in bilayer VBr$_2$.}
\textbf{a}, Top view of VBr$_2$. The black box denotes the magnetic supercell and the green dashed box denotes the primitive cell.
\textbf{b}, Side view of monolayer VBr$_2$.
\textbf{c}, Schematic magnetic configuration viewed along the in-plane [120] direction indicated by the red line in \textbf{a}, with black arrows marking the local magnetic moments on V sites.
\textbf{d}, Sliding ferroelectric switching from the AB to BA stacking through the intermediate AC stacking.
\textbf{e}, Stacking energy landscape as a function of in-plane fractional translation, with the energy of the AB (BA)-stacked states set to zero.
\textbf{f}, Transition energy barriers and out-of-plane polarization along the sliding ferroelectric switching path. 
\label{fig2}}
\end{figure}

\vspace{0.3cm}	
\noindent{\bf Sliding ferroelectricity in bilayer VBr$_2$}

\noindent
Based on symmetry considerations, we propose the two-dimensional van der Waals material VBr$_2$ as a representative realization of hybrid-parity sliding multiferroics. Bulk VBr$_2$ adopts the layered CdI$_2$-type structure, in which the magnetic moments on the V atoms form a triangular lattice. Geometrical frustration in the triangular lattice stabilizes a noncollinear but coplanar compensated magnetic order with a 120-deg angle between neighboring magnetic moments\cite{sodequistTypeIIMultiferroic2023,liuSpinChiralityDrivenMultiferroicityVan2024}. This magnetic structure has been experimentally confirmed by neutron-scattering measurements\cite{hirakawaStudyFrustrationEffects1983a,kadowakiNeutronScatteringStudy1985}. Furthermore, atomically thin VBr$_2$ has been successfully synthesized through chemical vapor deposition\cite{jiangGeneralSynthesis2D2023}. 


We begin with monolayer VBr$_2$, which provides the basis for understanding the structure and magnetism of its bilayer. As shown in Fig.~\ref{fig2}a,b, it adopts the 1T structure with space group $P\bar{3}m1$ (No.~164). The monolayer is centrosymmetric and therefore forbids electric polarization. The crystallographic primitive cell is indicated by the green frame in Fig.~\ref{fig2}a. The noncollinear 120-deg magnetic order triples the in-plane periodicity, resulting in a $\sqrt{3}\times\sqrt{3}\times1$ magnetic unit cell, as indicated by the black frame. The 120-deg magnetic state is illustrated in Fig.~\ref{fig2}c. The magnetic moments lie in the plane spanned by the crystallographic directions [120] and [001], taken as the $yz$ plane in our Cartesian coordinate system, and form a periodic noncollinear arrangement along the [120] direction. For comparison, we also consider the ferromagnetic state, the ferrimagnetic state, and the collinear antiferromagnetic state, as detailed in Supplementary Note 1. Our calculations show that the 120-deg magnetic state is the ground state, consistent with previous reports\cite{sodequistTypeIIMultiferroic2023,liuSpinChiralityDrivenMultiferroicityVan2024}.

For the bilayer, the top layer can be obtained by applying the operation $\{ C_{2y}{\parallel}M_{z}|\boldsymbol{t}+\boldsymbol{\tau}_{z}\} $ to the bottom layer, where $\boldsymbol{\tau}_{z}$ is a trivial out-of-plane translation and $\boldsymbol{t}$ is an in-plane fractional translation. The resulting bilayer possesses $\{ C_{2y}{\parallel}\tilde{M}_{z}\} $ symmetry in the absence of the in-plane translation, as detailed in Supplementary Note 1. Since the magnetic moments in the monolayer lie in the $yz$ plane, the bilayer preserves the coplanar magnetic structure. To identify the energetically favorable configuration with respect to the in-plane fractional translation $\boldsymbol{t}$, we calculate the stacking energy on a $6\times6$ grid of rigid in-plane translations. As shown in Fig.~\ref{fig2}e, the sliding energy landscape identifies the AB stacking with $\boldsymbol{t}=(1/3,0)$ and the BA stacking with $\boldsymbol{t}=(2/3,0)$ as the most stable configurations. 
The calculated in-plane lattice constant of the magnetic unit cell for the AB stacking is 6.506~{\AA}, corresponding to a lattice constant of 3.756~{\AA} for the crystallographic primitive cell, in good agreement with previously reported values for two-dimensional VBr$_2$\cite{sodequistTypeIIMultiferroic2023,liuSpinChiralityDrivenMultiferroicityVan2024}.


\begin{figure*}[t]
\centering
\includegraphics[width=1\textwidth]{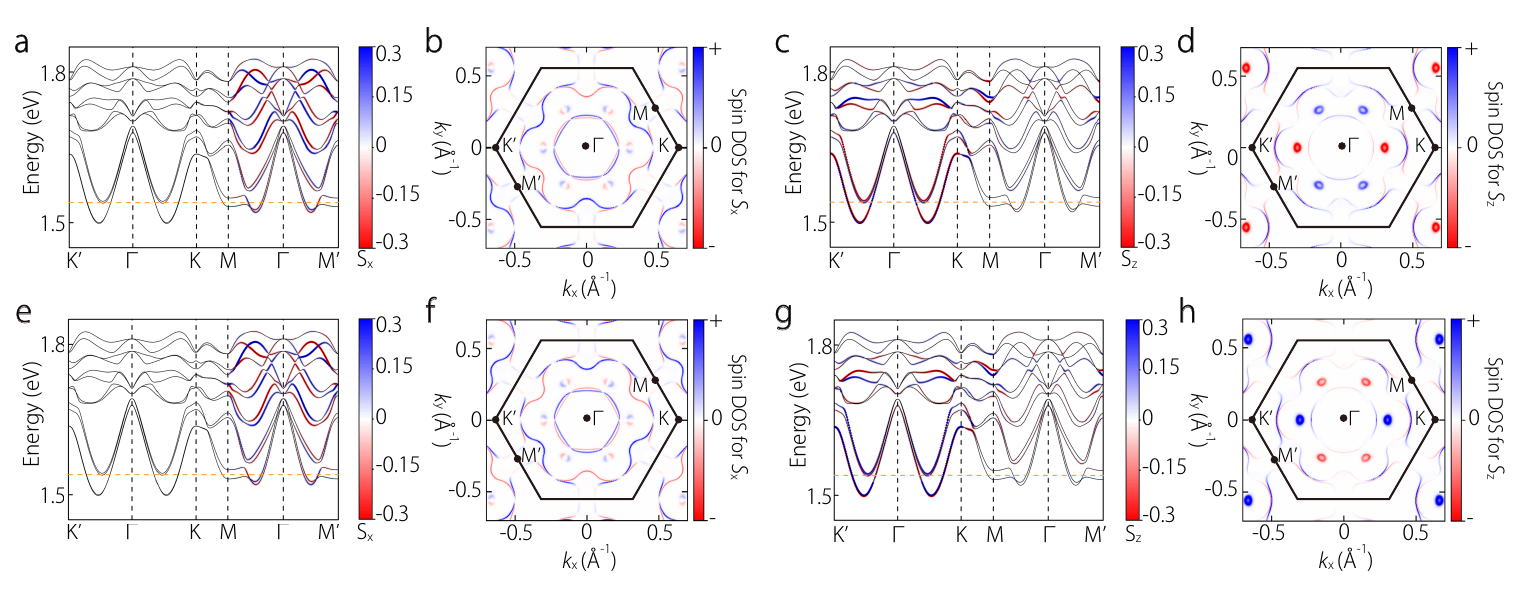}
\caption{\textbf{Switchable NSS components through interlayer sliding.}
\textbf{a}, The distribution of the spin expectation value $S_x$ in the band structure of the AB-stacked bilayer VBr$_2$. \textbf{b}, The spin density of states $\mathcal{N}_{S_x}$ for the AB-stacked bilayer on the Fermi surface contour at $E-E_{\mathrm{CBM}}=0.04$~eV, as indicated by the yellow dashed line in \textbf{a}. 
\textbf{c},\textbf{d}, The same as \textbf{a} and \textbf{b}, respectively, but for $S_z$ and the corresponding spin density of states $\mathcal{N}_{S_z}$ in the AB-stacked bilayer. 
\textbf{e}-\textbf{h}, The same as \textbf{a}-\textbf{d}, respectively, but for the BA-stacked bilayer. }
\label{fig3}
\end{figure*}

The result in Fig.~\ref{fig2}e also indicates that the AC stacking is the most favorable intermediate state along the sliding pathway. The AC-stacked bilayer corresponds to a nonpolar phase, which preserves $M_z$ symmetry. Using the climbing-image nudged elastic band method, we find that the energy barrier along this ferroelectric transition path is about 6~meV per formula unit, as shown in Fig.~\ref{fig2}f, indicating low-energy ferroelectric switching via interlayer sliding. The quotient SSG of the AB-stacked bilayer is generated by $\{C_{3x}^{-1}{\parallel}\tilde{C}_{3z}\}$ and $\{C_{2z}{\parallel}\tilde{M}_{y}\}$, which lacks $M_z$ symmetry and allows an out-of-plane electric polarization. Furthermore, since the AB- and BA-stacked bilayers are related by the operation $\{C_{2y}{\parallel}\tilde{M}_{z}\}$, the symmetry-allowed polarizations are opposite in the two stacking states, as shown in Fig.~\ref{fig2}d. The calculated out-of-plane polarization is shown in Fig.~\ref{fig2}f. The magnitude of the polarization reaches about 0.12~pC/m, which is comparable to those of experimentally verified sliding ferroelectrics, including bilayer $1T^{\prime}$-ReS$_2$ (0.07~pC/m)\cite{wanRoomTemperatureFerroelectricity12022} and bilayer WTe$_2$ (0.16~pC/m)\cite{feiFerroelectricSwitchingTwodimensional2018,wuSlidingFerroelectricity2D2021}.


The breaking of $M_z$ symmetry in the AB and BA stacking cases produces asymmetric interfacial coupling between layers. This asymmetry induces interlayer charge transfer, giving rise to opposite out-of-plane ferroelectric polarizations. To further clarify the interlayer charge transfer characteristics, we calculated the plane-averaged charge density difference along the $z$ direction together with the corresponding differential charge density maps for the AB and BA stacking configurations. As shown in Supplementary Fig.~3a,b, the charge accumulation is mainly concentrated in the interlayer region, forming a negative charge center, whereas charge depletion asymmetrically occurs within the two constituent layers. In the AB stacking, the charge depletion is stronger in the lower layer, placing the positive charge center below the negative charge center. Because the polarization points from the negative charge center toward the positive charge center, the AB stacking exhibits a spontaneous ferroelectric polarization along the $-z$ direction. After interlayer sliding to the BA stacking, the vertical charge redistribution is reversed, shifting the positive charge center above the negative charge center and giving rise to a ferroelectric polarization oriented along the $+z$ direction. This stacking-dependent reversal of the charge centers accounts for the opposite out-of-plane polarizations of the two ferroelectric states. The planar-averaged electrostatic potentials further exhibit potential steps of opposite signs across the AB and BA stackings, independently confirming the reversal of the out-of-plane ferroelectric polarization (see Supplementary Fig.~3c,d).

\vspace{0.3cm}	
\noindent{\bf Sliding-ferroelectric control of hybrid-parity NSS}

\noindent
Given the established ferroelectric switching between the AB and BA stacking configurations of bilayer VBr$_2$, we next investigate the simultaneously switched hybrid-parity NSS. Since bilayer VBr$_2$ preserves a coplanar magnetic structure lying in the $yz$ plane, its spin-only group is generated by $\{TC_{2x}{\parallel}1|\boldsymbol{0}\}$. This operation connects spin expectations at $\pm\boldsymbol{k}$, resulting in an odd parity for the $x$ component of NSS and even parity for the $y$ and $z$ components of NSS. Meanwhile, the AB- and BA-stacked bilayers are connected by $\{C_{2y}{\parallel}\tilde{M}_{z}\}$. It indicates that under the ferroelectric switching between the two stacking states, the $x$ and $z$ components of NSS are simultaneously switched, whereas the $y$ component of NSS remains unchanged (see Supplementary Fig.~4). Interestingly, bilayer VBr$_2$ enables the switching of both even- and odd-parity components of the NSS.

To confirm the hybrid-parity NSS and its switching under interlayer sliding, we evaluate the spin expectation distribution of the AB-stacked bilayer VBr$_{2}$ from first-principles calculations. The spin expectation value is calculated by $S_{p}\left(\boldsymbol{k}\right)=\bigl\langle u_{n\boldsymbol{k}}|\hat{s}_{p}|u_{n\boldsymbol{k}}\bigr\rangle$, where $\left|u_{n\boldsymbol{k}}\right\rangle $ is the Bloch eigenstate and $\hat{s}_{p}$ is the spin operator. The calculated distributions of the switchable components $S_{x}$ and $S_{z}$ are shown in Figs.~\ref{fig3}a and c, respectively. We further evaluate the spin distribution in the Brillouin zone at a given energy level $E$ by calculating the spin density of states $\mathcal{N}_{S_p}\left(E,\boldsymbol{k}\right)=\mathrm{Im}\mathrm{Tr}\bigl[G\left(E-i0^{+},\boldsymbol{k}\right)\hat{s}_p\bigr]/\pi$, where $G\left(E,\boldsymbol{k}\right)=(E-\hat{H}_{\boldsymbol{k}})^{-1}$ is the Green's function. Near the conduction band minimum, the calculated spin densities of states $\mathcal{N}_{S_x}$ and $\mathcal{N}_{S_z}$ are presented in Figs.~\ref{fig3}b and d, respectively. The NSS is clearly identified and becomes prominent near the Brillouin zone boundary. It is evident that the $x$ component of NSS exhibits odd parity, whereas the $z$ component exhibits even parity. The calculated spin distributions are also consistent with the quotient SSG of the AB-stacked bilayer. Specifically, the operation $\{C_{2z}{\parallel}\tilde{M}_{y}\}$ requires that under the mirror symmetry $M_y$, $S_x$ reverses its sign while $S_z$ remains unchanged. The operation $\{C_{3x}^{-1}{\parallel}\tilde{C}_{3z}\}$ requires $S_x$ to exhibit threefold rotation symmetry, whereas $S_z$ does not possess such symmetry. The results for the BA-stacked bilayer are presented in Figs.~\ref{fig3}e-h. The coexistence of odd- and even-parity NSS is also identified.  Moreover, both the odd-parity $x$ component and the even-parity $z$ component have opposite signs compared with those in the AB-stacked bilayer. More details of the hybrid-parity NSS in the AB- and BA-stacked bilayers are provided in Supplementary Figs. 5 and 6. These results are consistent with the symmetry analysis and confirm bilayer VBr$_2$ as a hybrid-parity sliding multiferroic.

\vspace{0.3cm}	
\noindent{\bf Signatures of hybrid-parity NSS switching}

\noindent
The NSS enables multiple intriguing phenomena in the absence of net magnetization. In collinear altermagnets, one typical transport effect is the magnetic spin current arising from NSS~\cite{GonzalezHernandez2021Efficient}. In this case, the spin current arises from the combined effects of even-parity NSS and the dipole modulation of the electron occupation induced by an external electric current. Here, we find that in hybrid-parity magnets, spin currents can be induced by either the even- or odd-parity components of NSS through the combination of current-driven dipole and quadrupole modulations of the electron occupation, respectively. This provides a probe of NSS in hybrid-parity magnets and can serve as a signature of hybrid-parity NSS switching.  

The spin currents arising from dipole and quadrupole modulations can be expressed as~\cite{hamamotoNonlinearSpinCurrent2017} $j^{\mathrm{D},p}_{a}=\sigma^{p}_{ab}E_{b}$ and $j^{\mathrm{Q},p}_{a}=\chi^{p}_{abc}E_{b}E_{c}$, where $\sigma^{p}_{ab}=-\int\partial_{E_{b}}f^{\mathrm{D}}v^{p}_{a}\left(\boldsymbol{k}\right)$ and $\chi^{p}_{abc}=-\int\partial_{E_{b}}\partial_{E_{c}}f^{\mathrm{Q}}v^{p}_{a}\left(\boldsymbol{k}\right)/2$. Here $v^{p}_{a}=\bigl\langle u_{n\boldsymbol{k}}|\{\hat{v}_{a},\hat{s}_{p}\}|u_{n\boldsymbol{k}}\bigr\rangle/2$ is the spin current of a given Bloch state, $\hat{v}_{a}$ is the velocity operator, $a$ denotes the current direction, $p$ denotes the spin direction of the spin current, $f^{\mathrm{D}}=\tau E_{b}\partial_{b}f_{0}$ and $f^{\mathrm{Q}}=\tau^2 E_{b}E_{c}\partial_{b}\partial_{c}f_{0}$ describe the dipole and quadrupole modulations of the electron occupation, respectively, $\tau$ is the relaxation time, and $f_{0}$ is the Fermi-Dirac distribution function. 
A key observation is that for systems with $\pm\boldsymbol{k}$ connected by symmetry, the dipole and quadrupole modulation terms possess definite parities. Specifically, the two distribution functions satisfy 
\begin{equation}
f^{\mathrm{D}}\left(\boldsymbol{k}\right)=-f^{\mathrm{D}}\left(-\boldsymbol{k}\right),\quad f^{\mathrm{Q}}\left(\boldsymbol{k}\right)=f^{\mathrm{Q}}\left(-\boldsymbol{k}\right).
\end{equation}
A spin current is allowed when the parity of the NSS is opposite to the parity of the distribution function. Therefore, one can immediately obtain that $j^{\mathrm{D}}$ and $j^{\mathrm{Q}}$ probe the even- and odd-parity components of NSS, respectively.

\begin{figure}[t]
\centering
\includegraphics[width=\columnwidth]{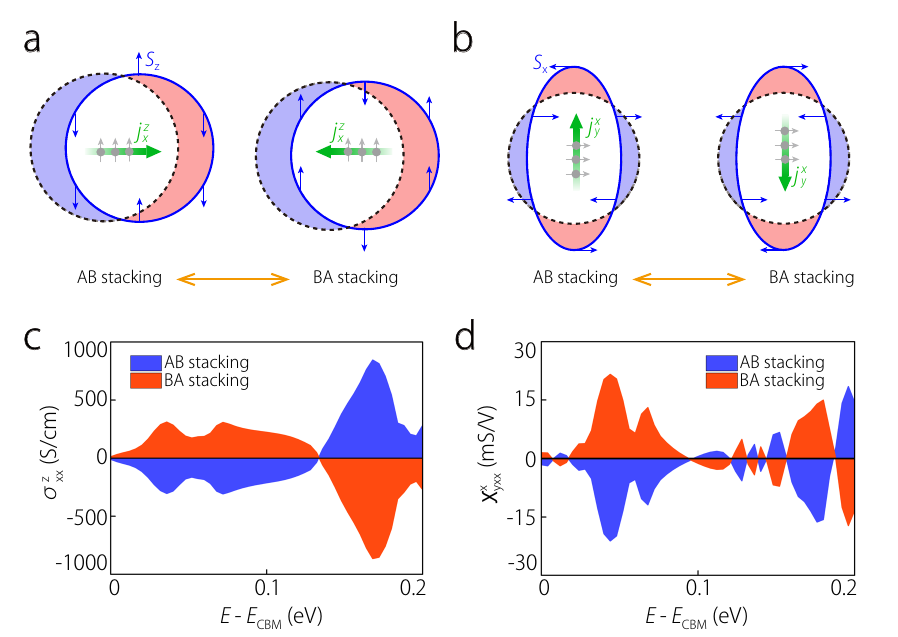}
\caption{\textbf{Signatures of hybrid-parity NSS switching in the bilayer VBr$_2$.}
\textbf{a}, \textbf{b}, Schematic illustrations of the generation of $j^{\mathrm{D}}$ and $j^{\mathrm{Q}}$, respectively. Each panel shows both the AB- and BA-stacked bilayer configurations. The blue arrows represent the even-parity $z$ component of the NSS in \textbf{a} and odd-parity $x$ component of the NSS in \textbf{b}. The dipole and quadrupole modulation of the electron occupation induced by the external driving field are shown in red and blue, representing positive and negative changes in occupation, respectively. The generated spin current $j^{p}_{a}$ is indicated by the green arrows, with the spin direction $p$ shown in grey. 
\textbf{c},\textbf{d}, Calculated $\sigma_{xx}^{z}$ and $\chi_{yxx}^{x}$ for AB- and BA- stacked bilayers.
\label{fig4}}
\end{figure}

For bilayer VBr$_{2}$, we perform a detailed symmetry analysis of the response functions $\sigma^{p}_{ab}$ and $\chi^{p}_{abc}$, as presented in Supplementary Note 2. We focus on the spin transport within the two-dimensional plane. By applying an electric field along the $x$ direction, the allowed spin currents are $j^{\mathrm{D},z}_{x}=\sigma^{z}_{xx}E_{x}$ and $j^{\mathrm{Q},x}_{y}=\chi^{x}_{yxx}E^{2}_{x}$, which correspond to a $z$-polarized spin current flowing along $x$ and an $x$-polarized spin current flowing along $y$, respectively. The even-parity $z$ component and the odd-parity $x$ component of the NSS are indeed responsible for $j^{\mathrm{D}}$ and $j^{\mathrm{Q}}$, respectively. The two flow along perpendicular directions, leading to spin accumulation polarized along different directions ($z$ and $x$) at different boundaries of the sample. Furthermore, we find that the operation $\{ C_{2y}{\parallel}\tilde{M}_{z}\} $ reverses both $j^{\mathrm{D},z}_{x}$ and $j^{\mathrm{Q},x}_{y}$. As shown in Fig.~\ref{fig4}a, for the even-parity $z$ component, the Bloch states at $\pm\boldsymbol{k}$ carry opposite spin currents. Combined with the dipole modulation of electron occupation, they generate a net spin current. Switching the NSS reverses the spin current carried by each Bloch state and consequently reverses the net spin current. A similar mechanism applies to the odd-parity $x$ component as shown in Fig.~\ref{fig4}b. The first-principles results of the response functions $\sigma^{z}_{xx}$ and $\chi^{x}_{yxx}$ for the AB-stacked bilayer in the absence of spin-orbit coupling are given in Figs.~\ref{fig4}c,d. Taking the relaxation time of 0.1~ps, the two response functions reach $\sigma^{z}_{xx}\sim$311~S/cm and $\chi^{x}_{yxx}\sim$22~mS/V near the conduction band minimum, respectively. The results of $\sigma^{z}_{xx}$ and $\chi^{x}_{yxx}$ with spin-orbit coupling are presented in Supplementary Fig. 7, showing that spin-orbit coupling makes a subdominant contribution to the two conductivities. The response functions for the BA-stacked bilayer are also calculated and are shown in Figs.~\ref{fig4}c,d. Both response functions exhibit opposite signs in the AB and BA stackings, consistent with the above analysis.

To summarize, we introduced multiferroicity in hybrid-parity magnets through sliding ferroelectricity, and showed that the hybrid-parity NSS in this compensated noncollinear magnet can be electrically controlled. Based on SSG symmetry analysis, we identified coplanar magnets as a natural platform for realizing this type of multiferroicity because their spin-only group symmetries enforce definite parities of NSS and are compatible with ferroelectricity. By introducing sliding ferroelectricity, the ferroelectric polarization and hybrid-parity NSS become coupled, resulting in hybrid-parity sliding multiferroics. Guided by symmetry considerations, bilayer VBr$_2$ was proposed as a representative example, and the switchable hybrid-parity NSS components were verified through first-principles calculations. We also showed that the spin currents serve as signatures of switching of both even- and odd-parity NSS components. The switchable even- and odd-parity NSS components in VBr$_2$ generate spin currents flowing along perpendicular directions. This leads to spin accumulation polarized along different directions at different boundaries, which can be detected through magneto-optical Kerr-effect measurements\cite{katoObservationSpinHall2004b,stammMagnetoOpticalDetectionSpin2017}. The spin currents can also be probed in nonlocal measurements, in which a magnetic electrode can be used to extract the spin-related signal\cite{songCoexistenceLargeConventional2020}. 
The tunneling magnetoresistance (TMR) has been proposed in altermagnets, arising from NSS\cite{smejkalGiantTunnelingMagnetoresistance2022a,shaoSpinneutralCurrentsSpintronics2021a}. More recent studies have also predicted a TMR effect in odd-parity magnets\cite{brekkeMinimalModelsTransport2024,yuanTimeReversalReserved2026b,ezawaTunnelingMagnetoresistanceJunction2026}. A similar scenario can be anticipated in hybrid-parity sliding multiferroics, where the magnetic configuration of the junction could be controlled by a gate field. Furthermore, altermagnets with even-parity NSS have shown considerable potential for spintronic applications, including spin field effect transistors\cite{zhuAltermagnetoelectricSpinField2025}, electrically switchable topological devices\cite{chenAltermagnetsEnableGateSwitchable2026}, spin filters\cite{fuAllelectricallyControlledSpintronics2025,samantaSpinFilteringInsulating2025a} and magnetic memory devices\cite{guoMagneticMemoryDriven2025}.  Given these developments, exploring spintronic applications based on electrically controllable hybrid-parity NSS represents an interesting direction for future work.

\vspace{0.3cm}
\noindent{\Large \bf Methods}

\vspace{0.05cm}	
\noindent{\bf First-principles calculations}

\noindent
The first-principles calculations based on density functional theory were performed using the Vienna \textit{ab initio} Simulation Package~\cite{Kresse1993Ab,Kresse1996Efficiency,Kresse1996Efficient}, using the projector-augmented wave method~\cite{Bloechl1994Projector}. The Perdew-Burke-Ernzerhof functional~\cite{Perdew1996Generalized} was adopted to address the exchange-correlation energy. The energy cutoff was set as 500~eV. The Monkhorst-Pack $k$-mesh with a size of $11\times 11\times 1$ was used for Brillouin zone sampling. The Dudarev \textit{et al.}\textquoteright s approach~\citep{Dudarev1998Electron} was used to treat localized $d$ orbitals of V atoms with an effective $U=1.0$~eV. The energy and force convergence criteria were set as 10$^{-6}$~eV and 0.01~eV/{\AA}, respectively. A vacuum layer with a thickness of about 26~{\AA} was taken to avoid artificial interactions between adjacent slabs. The interlayer van der Waals interaction was treated based on the DFT-D3 method of Grimme with zero-damping function~\cite{grimmeConsistentAccurateInitio2010,grimmeEffectDampingFunction2011}. The ferroelectric polarization was calculated using the Berry-phase method\cite{king-smithTheoryPolarizationCrystalline1993a}. The ferroelectric transition pathway was obtained by the climbing-image nudged elastic band method with seven intermediate images. All atomic coordinates of the intermediate images were optimized and the calculations were considered converged when the maximum force on any atom was below 0.03 eV/Å~\cite{henkelmanClimbingImageNudged2000,smidstrupImprovedInitialGuess2014}. To investigate the spin expectation, spin density of states and transport properties, the maximally localized Wannier functions were constructed by using Wannier90~\cite{Mostofi2014updated}. The $s$, $p$ and $d$ orbitals of V atoms and the $s$ and $p$ orbitals of Br atoms were used as the initial local basis. A dense $1000 \times 1000$ $k$-point mesh was used for Brillouin-zone integration to evaluate the conductivities. The plane-averaged charge density difference is defined as
\begin{equation}
\Delta \rho = \rho_{\mathrm{bilayer}}
- \rho_{\mathrm{upper}}
- \rho_{\mathrm{lower}},
\label{eq:charge_density_difference}
\end{equation}
where $\rho_{\mathrm{bilayer}}$, $\rho_{\mathrm{upper}}$, and
$\rho_{\mathrm{lower}}$ denote the charge densities of the bilayer,
upper layer, and lower layer of bilayer VBr$_2$, respectively.

\vspace{0.3cm}
\def\bibsection{\Large \noindent{\bf References}}	
\bibliographystyle{naturemag.bst}	
\bibliography{ref}

\vspace{0.3cm}
\noindent{\Large \bf Acknowledgements}

\noindent
This work is supported by the National Key R$\&$D Program of China (Grant No. 2022YFA1402600) (W.W.), the Australian Research Council Discovery Early Career Researcher Award (Grant No. DE240100627)  (X.W.), and the Australian Research Council Discovery Project (Grant No. DP260102992) (Z.C. and X.W.). Y.S.A. acknowledges financial support from the Singapore Ministry of Education (MOE) Academic Research Fund (AcRF) Tier 2 Grant No. MOE-T2EP50125-0019 (Y.S.A.) and the Kwan Im Thong Hood Cho Temple Early Career Chair Professorship (Y.S.A.). X.W. and Z.C. acknowledge the computational resources from the National Computational Infrastructure (NCI), which were allocated from the National Computational Merit Allocation Scheme supported by the Australian Government. 

The authors thank Dr. Shifeng Qian from Anhui Normal University, Dr. Qian Song from Massachusetts Institute of Technology, and Prof. Wei Wei from Shandong University for fruitful discussions.

\vspace{0.3cm}
\noindent{\Large \bf Author contributions}

\noindent
X.W. conceived and led the project and supervised the overall research. Z.G., J.C., and S.F. performed the symmetry analysis and first-principles calculations of the ferroelectric and magnetic properties. Z.G. and J.C. conducted the symmetry analysis and calculations of the transport properties. Z.G., S.F., J.C., Z.C. and X.W. prepared the manuscript with input from J.Y., X.Z., Y.S.A. and W.W..

\vspace{0.3cm}
\noindent{\Large  \bf Competing interests}
	
\noindent{The authors declare no competing interests.}
	
\vspace{0.3cm}
\noindent{\Large  \bf Additional information}
	
\noindent{{\bf Supplementary information} is available for this paper at https://doi.org/xxxxxx.}
	
\vspace{0.3cm}
\noindent{{\bf Reprints and permission information} is available at http://www.nature.com/reprints}
	
\vspace{0.3cm}
\noindent{{\bf Publisher's note} Springer Nature remains neutral with regard to jurisdictional claims in published maps and institutional affiliations.}

\end{document}